\documentclass[11pt]{article}
\usepackage{authblk}
\usepackage{amsmath}
\usepackage{graphicx}
\title{Angular Control of Anisotropy-Induced Bound States in the Continuum}

\author[1]{Samyobrata Mukherjee}
\author[1]{Jordi Gomis-Bresco}
\author[1]{Pilar Pujol-Closa}
\author[1,2,*]{David Artigas}
\author[1,2]{Lluis Torner}

\affil[1]{ICFO-Institut de Ciències Fotòniques, The Barcelona Institute of Science and Technology, 08860 Castelldefels (Barcelona), Spain}
\affil[2]{Department of Signal Theory and Communications, Universitat Politècnica de Catalunya, 08034 Barcelona, Spain}

\affil[*]{Corresponding author: david.artigas@icfo.eu}

\begin{document}
\maketitle

\begin{abstract}
	Radiation of leaky modes existing in anisotropic waveguides can be cancelled by destructive interference at special propagation directions relative to the optical axis orientation, resulting in fully bound states surrounded by radiative modes. Here we study the variation of the loci of such special directions in terms of the waveguide constitutive parameters. We show that the angular loci of the bound states are sensitive to several design parameters, allowing bound states to exist for a broad range of angular directions and wavelengths and suggesting applications in filtering and sensing.
\end{abstract}

\noindent Bound states in the continuum (BICs) are radiationless modes that exist in the part of the spectrum that corresponds to the continuum. Introduced by von Neumann and Wigner in quantum mechanics \cite{Neuman1929}, and later studied by Stillinger and Herrick \cite{Stillinger1975}, they have been demonstrated as a general wave phenomenon in acoustic systems \cite{Parker1966}, and more recently in different photonic systems \cite{Marinica2008, Bulgakov2008, Plotnik2011, Corrielli2013,Hsu2013,Monticone2017,Li2017, Minkov2018, Bulgakov2014,Bulgakov2015,Bulgakov2017a,Gomis-Bresco2017, Cerjan2019, Hsu2016}. Several practical applications of BICs have also been proposed \cite{Linton2007, Vallejo2010, Ramos2014, Monticone2014, Yoon2015, Rybin2017, Gansch2016, Kodigala2017, Midya2018, Kartashov2017, Romano2018}. Most BICs can be classified into symmetry/separability protected BICs and parameter tuning or interference BICs, depending on their underlying mechanism \cite{Hsu2013}. BICs belonging to the first class are sensitive to perturbations that break the symmetry, while BICs belonging to the second class may require suitable waveguide layouts, in the absence of which BICs couple to the continuum \cite{Rivera2016,Hu2018}. Topologically protected BICs are robust to variation of parameters \cite{Zhen2014, Bulgakov2017, Jin2018}.

\par Anisotropic planar waveguides containing uniaxial materials are known to support full-vector, hybrid modes, both fully guided and also semi-leaky \cite{Marcuse1979, Knoesen1988, Torner1993}. The semi-leaky modes are discrete, improper solutions of the eigenvalue equation with complex propagation constants whose imaginary part in most cases provides a good approximation to the actual radiation that is carried away by a continuous band of radiation modes \cite{Menyuk_2009}. The radiation channel is set by the polarization whose propagation eigenvalue is below cut-off, i.e., the ordinary or extraordinary polarizations for negative or positive birefringence, respectively. The leakage can be tailored \cite{Marcuse1979, Shipman2013}, and under suitable conditions even fully canceled, resulting in BICs \cite{Gomis-Bresco2017, Mukherjee2018, Timofeev2018}. Among anisotropy-induced BICs, polarization separable (PS) BICs, where the mode has a polarization orthogonal to the polarization of the radiation channel, are analogous to symmetry protected BICs. PS-BICs only exist in structures with anisotropy-symmetry, i.e., where the optical axes in all layers of the structure are aligned and parallel to the interfaces. Interference (INT) BICs are produced by careful design of the system such that the phase difference between the ordinary and extraordinary waves leads to destructive interference which cancels leakage into the radiation channel. They may exist for any orientation of the optical axis, are robust and show particularly rich physical features when the anisotropy-symmetry is broken \cite{Mukherjee2018}. 

\par PS-BICs are transversally polarized modes that are guided for optical axis orientations perpendicular to the propagation direction because they are affected only by the ordinary or the extraordinary index, with a higher index in the guiding core than in the claddings. Therefore PS-BICs behave as regular TE/TM modes of anisotropic waveguides and exist for all frequencies above cutoff. There is only one PS-BIC per semi-leaky branch \cite{Gomis-Bresco2017}. Changes in the refractive indices only affect the semi-leaky mode cutoff, but the properties of PS-BICs remain unaltered.

\par In this letter we therefore focus on anisotropy-induced INT-BICs and study the variation of their angular loci of existence and cut-off conditions when the main waveguide parameters vary. Specifically, we address variations in the refractive indices of the involved materials and in the waveguide thickness or in the operating wavelength. We show that the existence loci is sensitive to such parameters, which can be readily tuned in the design of the waveguide. Importantly, we also find that BIC lines may merge, creating large intervals of angular directions near a BIC at which light propagation may be virtually lossless.

\par
\begin{figure}[t]
    \centering
    \includegraphics[width=\linewidth]{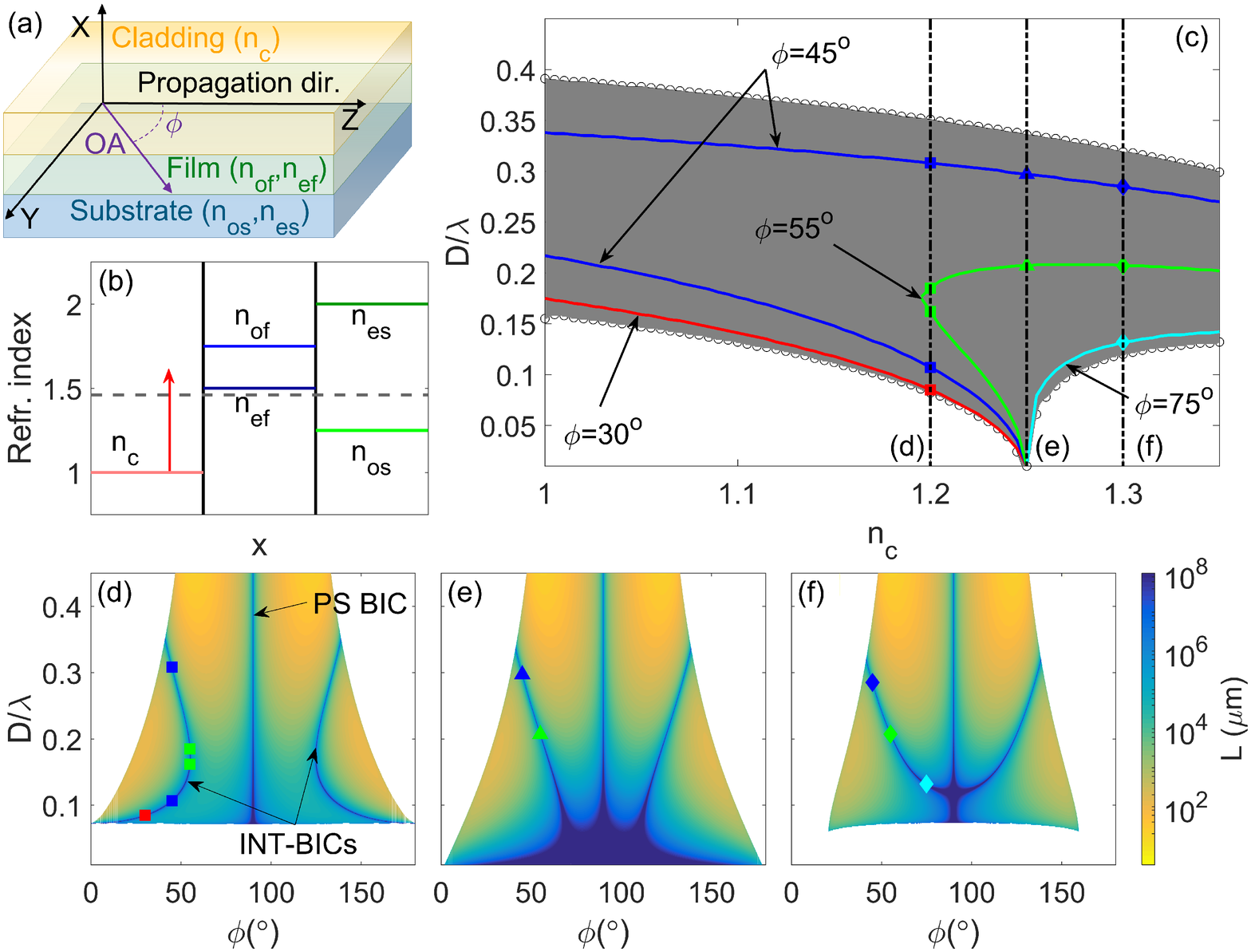}
    \caption{Impact of the variation of $n_c$ on the existence of INT-BICs. (a) Geometry of the structure. (b) Schematic of the refractive indices of the structure, with a film and substrate with negative ($n_{of}=1.75, n_{ef}=1.5$) and positive ($n_{os}=1.25, n_{es}=2$) birefringence, respectively. The red arrow shows that $n_c$ varies. The dashed grey line is an indication of the mode index $N$. (c) The colored lines show the loci of INT-BICs at specific values of $\phi$ in the $n_c$-$D/\lambda$ plane. The grey band shows the range of existence of the INT-BICs for all $\phi$. The existence loci of BICs in the $\phi$-$D/\lambda$ plane are shown for (d) $n_c=1.2$, (e) $n_c=1.25$ and (f) $n_c=1.3$, indicated by the three vertical dashed lines in (b). The color in (d)-(f) stands for the decay length of the leaky modes. BICs, with infinite decay length, appear as dark blue. The colored point-markers serve as visual guides for the BICs lines shown by the colored curves in (c).}
   % \\ \hrulefill}
    \label{fig:1}
\end{figure}

 We address anti-guiding planar waveguides to avoid the coexistence of BICs and guided modes. Then, either the ordinary or the extraordinary index in the substrate is the highest index in the structure, as depicted in Fig.\ref{fig:1}(b). The coordinate system is centered at the substrate/core interface, with $x$ being orthogonal to the interfaces and $z$ being the propagation direction. The electric fields in the substrate can be written in terms of ordinary and extraordinary waves as
\begin{equation}
     E_{is}= \left(A_{i,e} e^{-k_0 \gamma_{e} x} + A_{i,o} e^{- k_0\gamma_{o} x}\right) e^{- i N k_0 z}
     \label{fields}
\end{equation}
where $i=x, y$ or $z$ denotes the field component, $k_0$ is the free space wavenumber, $N$ the eigenmode effective index, and $D$ is the core thickness. The transverse decaying constant $\gamma_{e,o}$ for the evanescent ordinary and extraordinary wave are 
\begin{equation}
\begin{split}
    \gamma_o  = & \pm \sqrt{N^2-n_{os}^2}, \\
    \gamma_e  = & \pm \sqrt{N^2 \left((n_{es}^2/n_{os}^2) \cos^2{\phi}+\sin^2{\phi} \right)-n_{es}^2},
\end{split}
    \label{gammas}
\end{equation}
where $n_o/n_e$ refer to the ordinary and extraordinary refractive indices. The subscripts $s, f, c$ refer to the substrate, film and cover, respectively. $\phi$ refers to the angle between the orientation of the optical axes and the direction of propagation. Similarly, the field components in the guiding film can be written as
\begin{equation}
\begin{split}
    E_{if} & = (B_{1i,e} \sin{\left (k_0\kappa_{e} x \right )} + B_{1i,o} \sin{\left (k_0\kappa_{o} x \right )} \\
     & + B_{2i,e} \cos{\left (k_0\kappa_{e} x \right )} + B_{2i,o} \cos{\left (k_0\kappa_{o} x \right )}) e^{- i N k_0 z},
\end{split}
     \label{fieldf}
\end{equation}
where the transverse wavevector $\kappa_{e,o}$ for the ordinary and extraordinary field are
\begin{equation}
\begin{split}
    \kappa_o  = & \sqrt{n_{of}^2 - N^2}, \\
    \kappa_e  = &\sqrt{n_{ef}^2 - N^2 \left((n_{ef}^2/n_{of}^2) \cos^2{\phi}+\sin^2{\phi} \right)}.
\end{split}
\end{equation}
The fields in the cover can be written as
\begin{equation}
     E_{ic}= C_{ic} e^{k_0 \gamma_{c} \left(x + D \right)} e^{- i N k_0  z},
     \label{fieldc}
\end{equation}
with transverse decaying constant $\gamma_c$ for the evanescent field: 
\begin{equation}
    \gamma_c = \sqrt{N^2-n_{c}^2}.
\end{equation}
\indent The boundary conditions at the two interfaces, $x = 0, D$ yield the eigenmode equation (see Refs.~\cite{Gomis-Bresco2017, Mukherjee2018} for a detailed derivation), which can be formally written as
\begin{equation} 
    W \left(N,\phi, D/\lambda, \nu \right)=0.
    \label{trans}
\end{equation}
Here $\nu$ stands for all relevant refractive indices. Selecting the negative square root for $\gamma_e (\gamma_o)$ in a waveguide with a substrate with positive (negative) birefringence, yields the eigenvalue with complex $N$ for semi-leaky modes with a radiation channel associated to the extraordinary (ordinary) wave.

\par BICs appear when the field amplitudes of the radiation channel vanishes. Thus, setting $A_{ie}=0$ in (\ref{fields}) for a substrate with positive birrefringence, the resolution of the eigenmode problem yields the condition for BIC existence as
\begin{equation}
  M_{a} \kappa_{e} n_{of}^{2} \left(F \kappa_{o} C_o - \gamma_{o} S_o\right) + 
   \kappa_{o} \left(\gamma_{c} n_{c}^{2} S_e - Y \kappa_{e} n_{of}^{2} C_e\right)=0.
\label{auxe}
\end{equation}
Similarly, setting $A_{io}=0$ in (\ref{fields}) yields the condition for BIC existence in a structure with a substrate with negative birefringence,
\begin{equation}
  M_{a} \kappa_{o} \left(F \kappa_{e} n_{of}^{2} C_e - \gamma_{e} n_{os}^{2} S_e \right) +  \kappa_{e} n_{of}^{2} \left(\gamma_{c} S_o - Y \kappa_{o} C_o\right) = 0,
\label{auxo}
\end{equation}
where 
\begin{equation}
    \begin{split}
     S_{e(o)}= \sin( 2\pi\kappa_{e(o)} D/\lambda), & \quad C_{e(o)}= \cos( 2\pi\kappa_{e(o)} D/\lambda), \\
     F = \frac{N^{2} \cos^{2}{\left (\phi \right )}- n_{os}^{2},}{n_{of}^{2} - N^2 \cos^{2}{\left (\phi \right )}}, & \quad Y = \frac{N^{2}  \cos^{2}{\left (\phi \right )}- n_{c}^{2}}{n_{of}^{2} - N^2 \cos^{2}{\left (\phi \right )}}, \\ M_{a} = & \frac{(n_{of}^{2} - n_{c}^{2})}{(n_{of}^{2} - n_{os}^{2})}.
    \end{split}
    \label{mf}
\end{equation}

\par We first study a structure comprising a negative uniaxial core ($n_{of}>n_{ef}$) on a positive uniaxial substrate ($n_{os}<n_{es}$) with an isotropic dielectric as cover. The colored lines in Fig.~\ref{fig:1}(c) correspond to the loci of INT-BICs in the fundamental branch of the semi-leaky modes at a specific value of $\phi$ and the grey band shows the range of existence for INT-BICs for all values of $\phi$. The upper BIC cutoff, which decreases monotonically with $n_c$, corresponds to the transition from leaky to guided modes when $\gamma_e=0$. The lower BIC cutoff, decreasing up to the point where $n_c=n_{os}$, corresponds to the leaky mode cutoff to the substrate continuum of ordinary radiation waves, given by $\gamma_o=0$. At $n_c=n_{os}$ the structure becomes symmetric for the non-leaking polarization and like the fundamental guided mode in symmetric waveguides, the semi-leaky mode has no lower frequency cutoff. For $n_c>n_{os}$, the semi-leaky mode cutoff is dictated by the coupling to the continuum of the TE and TM radiation waves into the cover when $\gamma_c=0$. However, BICs now do not extend to the lower cutoff of the semi-leaky mode, and therefore the lower limit in the range of existence of INT-BICs in Fig.~\ref{fig:1}(c) corresponds to the minimum value of $D/\lambda$ at which INT-BICs exist. 

\par Figures \ref{fig:1}(d-f) show projections of the dispersion diagram in the $\phi$-$D/\lambda$ plane. Losses for the semi-leaky branch are indicated by the length $L(\mu m)$, defined as the length at which the amplitude of the semi-leaky mode is reduced to $1/e$ (in the color scale). The blue lines with the highest values of $L$ are given by (\ref{auxe}), and therefore correspond to BICs. The central blue line at $\phi=90^{\circ}$ in Figs.~\ref{fig:1}(d-f) corresponds to the aforementioned pure TM, PS-BIC, which exists for all values of $D/\lambda$ above the semi-leaky mode cutoff. As the waveguide is anisotropy-symmetric, the distribution of the INT-BICs is symmetric around $\phi=90^{\circ}$. For sufficiently small values of $D/\lambda$ and $n_c=n_{os}$, the condition for BIC existence (\ref{auxe}) is always fulfilled for any propagation direction $\phi$.  As $n_{c}$ increases, the lower part of the BIC lines approach $\phi=90^{\circ}$, and at $n_c=n_{os}$, the lines of INT-BICs merge via the BICs at small values of $D/\lambda$ resulting in a region of negligible losses on the semi-leaky mode. As $n_c$ increases further, with $n_c>n_{os}$, the lower cutoff of the INT-BICs is given by the points where the INT-BIC lines merge with the PS BIC, which still shows a region of very low losses near the merging point. The locus of this point fulfills the equation that is obtained by combining (\ref{trans}) and (\ref{auxe}) at $\phi=90^{\circ}$, which writes:
\begin{equation}
    M_a \kappa_e \left(n^2_{os} \kappa_o C_o + n^2_{of} \gamma_o S_o\right) - \kappa_o n^2_c \left(\gamma_c S_e + \kappa_e C_e \right)=0.
    \label{merge_eq_1}
\end{equation}

\begin{figure}[t]
    \centering
    \includegraphics[width=0.9 \linewidth]{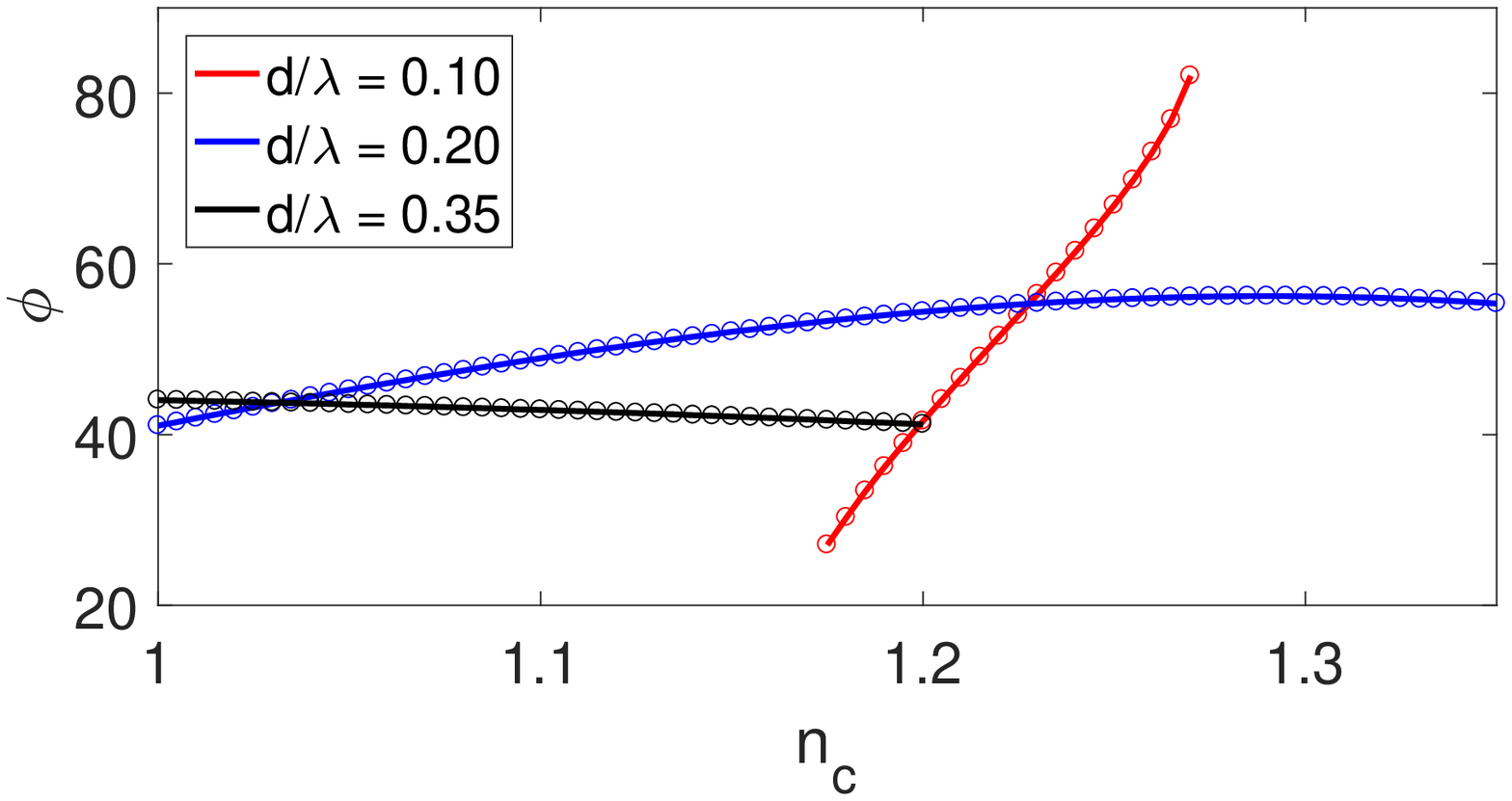}
    \caption{Loci of existence of the INT-BICs as a function of the refractive index of the cover $n_c$, for the waveguide in Fig.~\ref{fig:1}}
    %\\ \hrulefill}
    \label{fig:2}
\end{figure}
Figure \ref{fig:2} shows the variation of the angular loci of existence of INT-BICs relative to optical axis orientation $\phi$, in the same structure but for three different values of $D/\lambda$, as a function of the isotropic core refractive index $n_c$, in the range $0^{\circ}<\phi<90^{\circ}$. The plot shows that a variety of sensitivities and tendencies in $\phi$ are possible by choosing the proper value of $D/\lambda$. The highest sensitivity is obtained at low values of $D/\lambda$, where, e.g., a change in $n_c$ from $1.175$ to $1.27$ corresponds to a monotonic change in $\phi$ from  $27^{\circ}$ to $82^{\circ}$, for $D/\lambda=0.1$. The range of values of $n_c$ for which INT-BICs exist depends on the actual values of ${n_{of}, n_{ef}}$ and ${n_{os}, n_{es}}$ and is given by the lower BIC cutoff in Fig.~\ref{fig:1}(c). For $D/\lambda=0.2$, the sensitivity of the angular loci of existence of the BICs as a function of $n_c$ is lost. However, BICs exist for any value of $n_c$, since its existence is not affected by the lower cutoff in Fig.~\ref{fig:1}(c). At an even higher value of $D/\lambda$, i.e., $D/\lambda=0.35$, the sensitivity remains low, but now the BIC reaches the upper cutoff in Fig.~\ref{fig:1}(c) at $n_c=1.2$. 

The above results show that INT-BICs show promise for sensing devices where mode losses change under variation of extrinsic parameters, such as the cover refractive index \cite{Romano2018}. Fig.~\ref{fig:2} shows that the lower the value of $D/\lambda$ the higher the sensitivity. Also, the sensitivity of the INT-BICs to the direction of propagation relative to the optic axis orientation suggests their use as spatial angular filters as only light propagating along the direction around a BIC propagates with low-losses \cite{Takayama2014}.

\begin{figure}[t!]
    \centering
    \includegraphics[width=\linewidth]{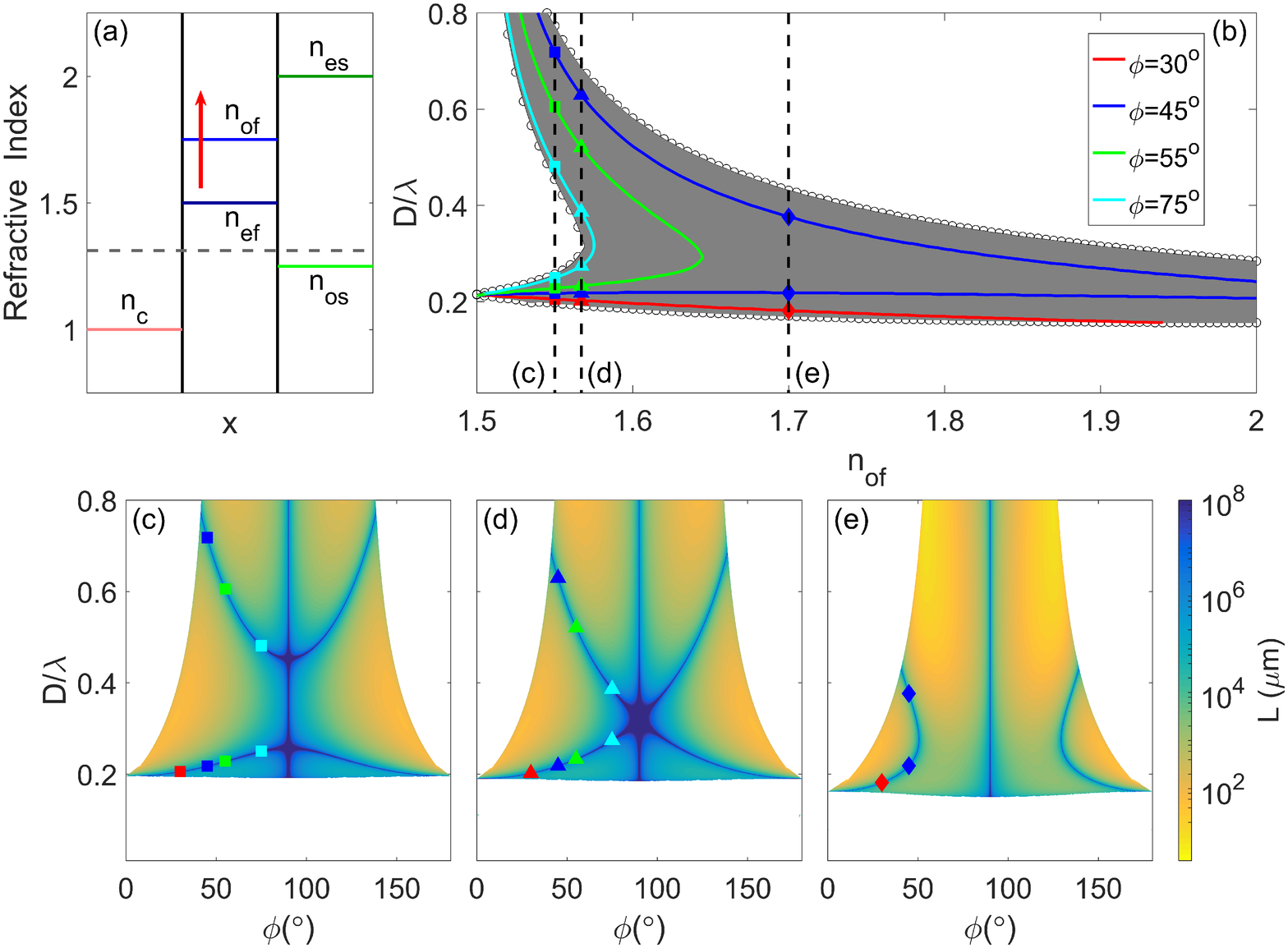}
    \caption{Impact of the variation of $n_{of}$ on the existence of INT-BICs. (a) Schematic of the refractive indices of the structure. The red arrow shows the parameter being varied and the dashed grey line indicates the mode index. The other parameters are the same as in Fig.~\ref{fig:1}, but now with $n_c=1$. (b) $D/\lambda$  range of existence of INT-BICs as a function of $n_{of}$. The colored lines show the loci of INT-BICs at specific values of $\phi$. Existence loci of BICs in the $\phi - D/\lambda$ plane, for (c) $n_{of}=1.55$, (d) $n_{of}=1.567$ and (e) $n_{of}=1.7$. The colored point-markers serve as visual guides for the BICs lines shown by the colored curves in (b).}
    %\\ \hrulefill}
    \label{fig:3}
\end{figure}

Figure \ref{fig:3} shows the impact of the variation of $n_{of}$ on the existence loci of INT-BICs for a structure with the same parameters as in Fig.~\ref{fig:1}, but now for $n_c=1$. When $n_{of}=n_{ef}$, the core is an isotropic material, thus existence of INT-BICs is  no longer possible and only PS-BICs may exist \cite{Gomis-Bresco2017}. When $n_{of}>n_{ef}$, INT-BICs are possible and their existence loci bends and forms two bands in \ref{fig:3}(b) one starting at the lower $D/\lambda$ cutoff, and one approaching from $D/\lambda >>1$, as $n_{of}$ increases. An example of a dispersion diagram in this region is shown in Fig.~\ref{fig:3}(c) for $n_{of}=1.55$, where a lower BIC line is visible, ascending from the lower cutoff, while an upper BIC line is seen approaching from infinity. When $n_{of}=1.567$, all BIC lines merge at $\phi=90^{\circ}$ and $D/\lambda=0.325$, as shown in Fig.~\ref{fig:3}(d), resulting in large region of very low loss in the semi-leaky modes around the point of intersection of the BIC lines. The locus of this point can be obtained by simultaneously solving (\ref{trans}, \ref{merge_eq_1}). The existence of a region at which the leaky mode exhibits very small propagation losses is important for practical applications \cite{Jin2018} such as a band pass filter in both wavelength and direction. Upon increasing $n_{of}$ further, the lines of INT-BICs stop intersecting with the PS-BIC and separate in $\phi$ as shown in Fig.~\ref{fig:3}(e) for $n_{of}=1.7$. 

\begin{figure}[t]
    \centering
    \includegraphics[width=\linewidth]{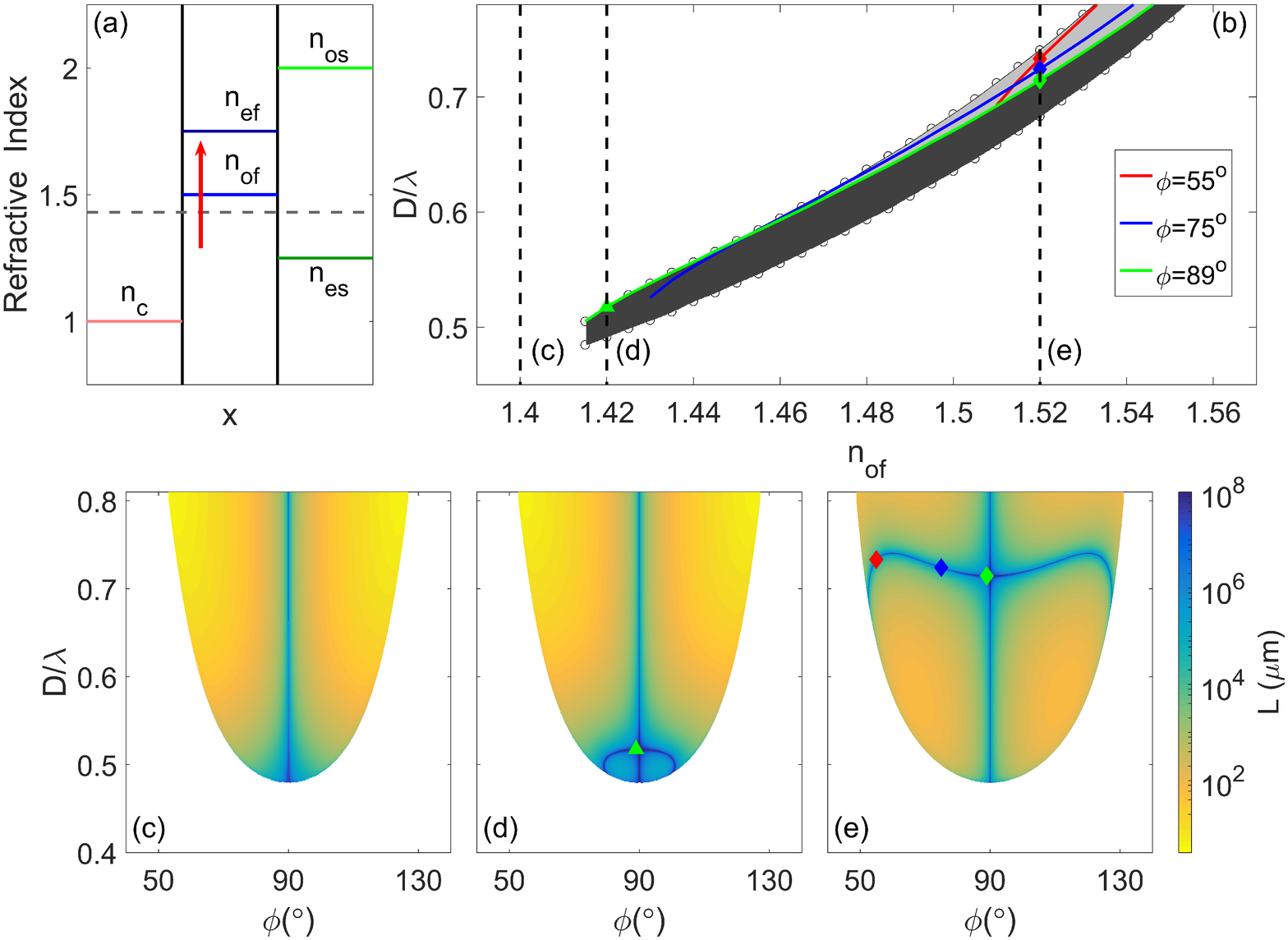}
    \caption{Impact of the variation of $n_{of}$ on the existence of INT-BICs in a structure with (a) isotropic core ($n_c=1$), and positive ($n_{ef}=1.75$) and negative ($n_{os}=2, n_{es}=1.25$) birefringent core and substrate, respectively. The red arrow shows the parameter being varied and the dashed grey line indicates the mode index. (b) $D/\lambda$ range of existence of INT-BICs in the second semi-leaky branch as a function of $n_{of}$. The colored lines show the loci of INT-BICs at specific values of $\phi$. Existence loci of BICs in the $\phi - D/\lambda$ plane for (c)  $n_{of}=1.4$, (d) $n_{of}=1.42$ and (e) $n_{of}=1.52$. The colored point-markers serve as visual guides for the BICs lines shown by the colored curves in (b). }
    %\\ \hrulefill}
    \label{fig:4}
\end{figure}

Figure \ref{fig:4} corresponds to a structure with a positive birefringent core and a negative birefringent substrate, when $n_{of}$ is varied. INT-BICs in this structure first appear on the second semi-leaky mode. The range of $D/\lambda$ at which the INT-BICs exist, Fig.~\ref{fig:4}(b), shows only one band appearing above the value $n^c_{of}=1.415$. Below $n^c_{of}$, only PS-BICs exist [Fig.~\ref{fig:4}(c)]. Just above $n^c_{of}$, a line of INT-BICs appears at the lower frequency cutoff [Fig.~\ref{fig:4}(d)], and evolves to higher values of $D/\lambda$ as $n_{of}$ increases, as shown in Fig.~\ref{fig:4}(e). This results in a band of BICs that ascends in $D/\lambda$ until $n_{of}=n_{ef}$, the point at which INT-BICs exist at $D/\lambda >>1$. At low values of $n_{of}$, the line of INT-BICs on the semi-leaky mode increases monotonically in $D/\lambda$ within the range $0^{\circ} < \phi <90^{\circ}$, and as a consequence, at a given value of $D/\lambda$ there is only one INT-BIC [Fig.~\ref{fig:4}(d)]. At higher values of $n_{of}$, the line of existence of INT-BICs is not monotonic, and therefore this can result in more than one BIC propagating in different directions, for a given value of $D/\lambda$  [Fig.~\ref{fig:4}(e)]. This situation is shown in Fig.~\ref{fig:4}(b), where the dark and light grey colour correspond to one and two INT-BICs, respectively, in the range $0^{\circ}< \phi < 90^{\circ}$. For this structure, the INT-BIC existence lines always intersect the line of existence of PS-BICs.  The locus of this point fulfills the equation that is obtained by combining (\ref{trans}) and (\ref{auxo}) at $\phi=90^{\circ}$, which writes:
\begin{equation}
    M_a \kappa_o n^2_{os} \left(\kappa_e C_e + \gamma_e S_e \right)- \kappa_e \left(n^2_{of} \gamma_c S_o + n^2_c \kappa_o C_o \right)=0.
    \label{merge_eq_2}
\end{equation}
In summary, we have found the loci of existence and cut-off conditions for anisotropy-induced BICs and have shown that they can be readily tuned by varying the various waveguide parameters involved, namely the material refractive indices, the core thickness and the operating wavelength. The lines of existence of the interference BICs have been found to intersect at certain parameter values, leading to areas where low-loss propagation is possible around the BICs. These findings enhance the understanding of full-vector BICs in anisotropic waveguides and suggest applications to devices based on the angular control of the propagation losses of light in integrated optics structures. \\

\par This work was supported by the Generalitat de Catalunya, CERCA, AGAUR 2017-SGR-1400; the Govt. of Spain through grant PGC2018-097035-B-I00, Severo Ochoa 2016-2019 grant SEV-2015-0522; Fundaci\'{o} Cellex; Fundaci\'{o} Mir-Puig. This project has received funding from the European Union's Horizon 2020 research and innovation programme under the Marie Sk\l odowska-Curie grant agreement No 665884.

%\bigskip
% Bibliography
\bibliography{bibliography}
\bibliographystyle{ieeetr}

% Full bibliography added automatically for Optics Letters submissions; the following line will simply be ignored if submitting to other journals.
% Note that this extra page will not count against page length
%\bibliographyfullrefs{bibliography}

\end{document}